\documentclass[11pt,english]{article}
\usepackage{axodraw}
\usepackage{latexsym}
\usepackage{graphicx}
\usepackage{psfrag}
\usepackage{amsmath,amssymb}
\def\ltsim{\lower3pt\hbox{$\, \buildrel < \over \sim \, $}}  
\def\gtsim{\lower3pt\hbox{$\, \buildrel > \over \sim \, $}}  
\makeatletter
\def\section{\@startsection {section}{1}{\z@}{-3.5ex plus -1ex minus
 -.2ex}{2.3ex plus .2ex}{\large\bf}}
\def\subsection{\@startsection{subsection}{2}{\z@}{-3.25ex plus -1ex
minus -.2ex}{1.5ex plus .2ex}{\normalsize\bf}}
\makeatother
\newcommand{\captionfonts}{\small}
\makeatletter  % Allow the use of @ in command names
\long\def\@makecaption#1#2{%
  \vskip\abovecaptionskip
  \sbox\@tempboxa{{\captionfonts #1: #2}}%
  \ifdim \wd\@tempboxa >\hsize
    {\captionfonts #1: #2\par}
   \else
    \hbox to\hsize{\hfil\box\@tempboxa\hfil}%
  \fi
  \vskip\belowcaptionskip}
\makeatother   % Cancel the effect of \makeatletter
\catcode`@=11
\def\marginnote#1{}
\newcount\hour
\newcount\minute
\newtoks\amorpm
\hour=\time\divide\hour
by60
\minute=\time{\multiply\hour by60 \global\advance\minute
by-\hour}
\edef\standardtime{{\ifnum\hour<12 \global\amorpm={am}
\else\global\amorpm={pm}\advance\hour by-12 \fi
 \ifnum\hour=0
\hour=12 \fi
 \number\hour:\ifnum\minute<10
0\fi\number\minute\the\amorpm}}
\edef\militarytime{\number\hour:\ifnum\minute<10
0\fi\number\minute}
\def\draftlabel#1{{\@bsphack\if@filesw
{\let\thepage\relax
 \xdef\@gtempa{\write\@auxout{\string
\newlabel{#1}{{\@currentlabel}{\thepage}}}}}\@gtempa
 \if@nobreak
\ifvmode\nobreak\fi\fi\fi\@esphack}
\gdef\@eqnlabel{#1}}
\def\@eqnlabel{}
\def\@vacuum{}
\def\draftmarginnote#1{\marginpar{\raggedright\scriptsize\tt#1}}
\def\draft{\oddsidemargin
0.0truein
 \def\@oddfoot{\sl preliminary draft \hfil
\rm\thepage\hfil\sl\today\quad\militarytime}
 \let\@evenfoot\@oddfoot
\overfullrule 3pt
 \let\label=\draftlabel
\let\marginnote=\draftmarginnote
\def\@eqnnum{(\theequation)\rlap{\kern\marginparsep\tt\@eqnlabel}
\global\let\@eqnlabel\@vacuum}
}
\catcode`@=12
\def\XXint#1#2#3{{\setbox0=\hbox{$#1{#2#3}{\int}$}
     \vcenter{\hbox{$#2#3$}}\kern-.5\wd0}}

\def\bea{\begin{eqnarray}} \def\eea{\end{eqnarray}}
\def\be{\begin{eqnarray}} \def\ee{\end{eqnarray}} 
\def\lsim{\mathrel{\rlap{\lower4pt\hbox{\hskip1pt$\sim$}}
    \raise1pt\hbox{$<$}}}                % less than or approx. symbol
\def\gsim{\mathrel{\rlap{\lower4pt\hbox{\hskip1pt$\sim$}}
    \raise1pt\hbox{$>$}}}                % greater than or approx. symbol

\newcommand{\promille}{%
  \relax\ifmmode\promillezeichen
        \else\leavevmode\(\mathsurround=0pt\promillezeichen\)\fi}
\newcommand{\promillezeichen}{%
  \kern-.05em%
  \raise.5ex\hbox{\the\scriptfont0 0}%
  \kern-.15em/\kern-.15em%
  \lower.25ex\hbox{\the\scriptfont0 00}}
\begin{document}

\thispagestyle{empty}

\begin{center}
\hfill CERN-PH-TH/2008-251\\
\hfill UAB-FT-659

\begin{center}

\vspace{1.7cm}

{\LARGE\bf Conformal Neutrinos: \\ \vspace{.5cm}
an Alternative to the See-saw Mechanism}

\end{center}

\vspace{1.4cm}

{\bf Gero von Gersdorff$^{\,a}$ and Mariano Quir\'os$^{\,b}$}\\

\vspace{1.2cm}

${}^a\!\!$ {\em {CERN Theory Division, CH-1211 Geneva 23, Switzerland}}

${}^b\!\!$
{\em { IFAE, Universitat Aut{\`o}noma de Barcelona,
08193 Bellaterra, Barcelona (Spain)}}

{\em {and}}

{\em {Instituci\`o Catalana de Recerca i Estudis  Avan\c{c}ats (ICREA)}}

\end{center}

\vspace{0.8cm}

\centerline{\bf Abstract}
\vspace{2 mm}
\begin{quote}\small
We analyze a scenario where the right-handed neutrinos make part of a
strongly coupled conformal field theory and acquire an anomalous
dimension $\gamma<1$ at a large scale $\Lambda$. Their Yukawa
couplings to the Higgs become irrelevant at the fixed point and they
are suppressed at low scales giving rise naturally to a small
(sub-meV) Dirac neutrino mass which breaks the conformal invariance.
We derive an upper bound on $\gamma$ from loop-induced flavor changing
neutral currents. Neutrino Yukawa couplings can be sizable at
electroweak scales and therefore the invisible decay of the Higgs in
the neutrino channel can be comparable to the $c\bar c$ and
$\tau\bar\tau$ modes and predict interesting Higgs phenomenology.  If
lepton number is violated in the conformal theory an irrelevant
Majorana mass operator for right-handed neutrinos appears for
$\gamma>1/2$ giving rise to an inverse see-saw mechanism. In this case
light sterile neutrinos do appear and neutrino oscillation experiments
are able to probe our model.
\end{quote}

\vfill

\newpage

It is by now a well established experimental fact that neutrinos have
nonzero masses.  On the other hand the nature and absolute scale of
these masses are far from clear. The most important theoretical
implication from this experimental evidence is that there have to
exist right-handed neutrino fields, as direct (Majorana) mass terms
for left-handed neutrinos are forbidden by the electroweak gauge
symmetry of the Standard Model (SM). It is then possible to write down
Yukawa couplings of the neutrinos to the Higgs field, thereby
generating Dirac mass terms ($m_\nu^D$) after electroweak symmetry
breaking (EWSB) in the standard way. On top of that one can allow for
Majorana mass terms for the right-handed neutrinos as the latter are
totally sterile with respect to the SM. The most successful model so
far consists of assigning to the right-handed neutrino a huge Majorana
mass ($m_{\nu_R}^M$), possibly generated by the breaking of a Grand
Unified Theory (GUT) at a scale of the order $10^{16}$ GeV. Sub-eV
neutrino masses are then achieved via the so-called see-saw
mechanism~\cite{Minkowski:1977sc} by noting that the light eigenvalue
of this system is given approximately by
\be
m_{\nu}\simeq \frac{(m_\nu^D)^2}{m_{\nu_R}^M}\,.
\ee
This is a very neat way to explain the smallness of the neutrino
masses without resorting to unnaturally small Yukawa couplings. The
see-saw mechanism then implies that neutrinos are Majorana particles,
a fact that can be tested experimentally by observing neutrinoless
double beta (0$\nu\beta\beta$) decay of certain nuclei. The only
drawback of this otherwise beautiful and simple mechanism is that it
involves either physics at an energy scale inaccessible to present and
future high energy colliders or tuning the Yukawa couplings to
extremely small values.  It was further suggested in
Refs.~\cite{Georgi:1983mq,Nelson:2000sn} that small Yukawa couplings
can be naturally achieved by assuming that physics right below the GUT
scale is governed by a strongly coupled infrared (IR) fixed point that
results in positive anomalous dimensions for the matter fields such
that Yukawa couplings become irrelevant operators. Conformal symmetry
is broken at some intermediate scale $M_{\rm W}\ll M_{\rm int}\ll
M_{\rm GUT}$ and Yukawa hierarchies can be generated in a natural
way~\footnote{Certain supersymmetric models also allow for a
suppression of flavor dependence of the soft
masses~\cite{Nelson:2000sn,Kobayashi:2001kz,Nelson:2001mq,Luty:2001jh},
a mechanism that is sometimes referred to as conformal
sequestering. Furthermore models with dynamical symmetry breaking
involving quasi-conformal behavior and large anomalous dimensions have
also been proposed in Ref.~\cite{Appelquist:2002me}.  }.

In this letter we would like to report on an alternative and natural
way to obtain small neutrino masses.  We will assume that only the
right-handed neutrinos $\mathcal N_R$ make part of a conformal theory
with a fixed point at the scale $\Lambda$: in the language of
Ref.~\cite{Georgi} right-handed neutrinos are unparticles with large
anomalous dimension that couple to SM fields through irrelevant
operators. Moreover if lepton number is not conserved in the conformal
theory irrelevant Majorana mass terms for right-handed neutrinos are
present, together with dimension-five couplings between the leptons
and Higgs. This can yield interesting modifications of our mechanism.
Conformal symmetry breaking will be induced by the electroweak
breaking at the scale of the Dirac neutrino mass while all the
neutrino phenomenology [as e.g.~$\mu\to e\gamma$ and other flavor
changing rare processes, or $h\to\nu\bar\nu$] will take place at
scales where the right-handed neutrinos keep their unparticle nature
giving rise to interesting and new phenomena.

Our four-dimensional (unparticle-like) approach should have a
five-dimensional counterpart where the conformal invariance is broken
by a mass gap~\cite{Cacciapaglia:2008ns}. It is essentially different
to higher-dimensional theories where right-handed neutrinos propagate
in a five-dimensional space and conformal invariance is broken by an
IR brane~\cite{Xdim}.

If the theory is strongly coupled the field $\mathcal N_R$ may acquire
a large anomalous dimension $\gamma$ and its propagator in
two-component spinor notation is given by~\cite{Cacciapaglia:2008ns}
\begin{equation}
\Delta(p,\gamma)
%\langle\nu_R\bar\nu_R\rangle
=-i B_\gamma \bar\sigma^\mu p_\mu
(-p^2-i\epsilon)^{-1+\gamma}, \quad
B_\gamma=
\frac{\Gamma(1-\gamma)}{
(4\pi)^{2\gamma}\,\Gamma(1+\gamma)}\,,
%\frac{8\pi^{3/2}}{(2\pi)^{2+2\gamma}}
%\frac{\Gamma(1-\gamma)\Gamma(3/2+\gamma)}{\Gamma(2+2\gamma)}\,,
\label{prop}
\end{equation}
where the particle limit is reached for $\gamma=0$ and $B_0=1$. For
$\gamma>0$, the renormalizable operator with the Standard Model fields
$\bar\ell_L H \mathcal N_R$ (where $H$ is the Standard Model Higgs doublet and
$\ell$ the leptonic doublet) becomes irrelevant. 
%In the same way the
%neutrino Majorana mass $\mathcal N_R \mathcal N_R$ also becomes irrelevant if
%$\gamma>1/2$

We will now assume that the UV theory conserves lepton number and 
come back to the effect of Majorana mass terms at the end of the paper.
The effective Lagrangian at the scale $\Lambda$ is then given by
\begin{equation}
\mathcal L(\Lambda)=\Lambda^{-\gamma}\, \ell_L H
\bar{\mathcal N}_R
%+\frac{1}{2}\Lambda^{1-2\gamma}\, \mathcal N_R \mathcal N_R 
+ h.c.
\label{lagrangiano}
\end{equation}
where we are fixing the Yukawa coupling at the $\Lambda$ scale as
$h_\nu(\Lambda)=1$.
The fact that the Yulawa coupling in Eq.~(\ref{lagrangiano}) is sequestered
by the conformal dynamics for scales $\mu<\Lambda$ can be made
explicit by redefining $\mathcal N_R$ in terms of fields $\nu_R$ with canonical
dimension as
\begin{equation}
\mathcal N_R= B_\gamma^{1/2}\,\mu^\gamma\,\nu_R\,.
\label{relacion}
\end{equation}
Therefore for scales $\mu<\Lambda$ one can write the effective Lagrangian
\begin{equation}
\mathcal
L(\mu)=B_\gamma^{1/2}\left(\frac{\mu}{\Lambda}\right)^{\gamma}\,\ell_L
H
\bar\nu_R + h.c.
\label{lagrangiano2}
\end{equation}
The coefficient $B_\gamma$ varies very little over the values of
$\gamma$ considered here. Although it diverges as $(1-\gamma)^{-1}$
for $\gamma$ close to one, it does so with a small coefficient and one
has only a mild variation $B_\gamma\approx 0.09-0.16$ for $\gamma$ in
the range $0.5-0.95$.

\begin{center}
\begin{figure}[thb]
\psfrag{logLambda}{$\log_{10}(\Lambda/$GeV)}
\psfrag{g}[][t]{$\gamma$}
\psfrag{r}[][t]{$\log_{10}(m_{\nu_R}^M/m_\nu^D)$}
\centering
\includegraphics[height=5cm]{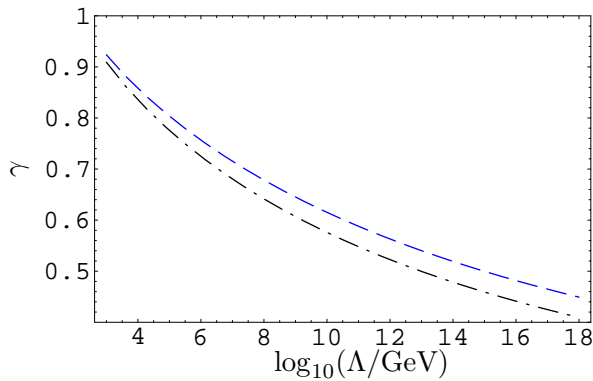}
\caption{The anomalous dimension $\gamma$ of the right-handed neutrino
as a function of the scale $\Lambda$. The two lines correspond to
neutrino masses of $m^D_\nu=0.01$ eV (dashed blue line) and $1$ eV
(dash-dotted black line).}
\label{fig1}
\end{figure}
\end{center}

When the Higgs field acquires a vacuum expectation value,
%as a consequence of the Standard Model dynamics
 $\langle H\rangle=v/\sqrt{2}$, the resulting Dirac mass term
represents a tiny relevant perturbation to the conformal sector that
will eventually drive it away from the fixed point. We thus face the
intriguing possibility that electroweak breaking itself is responsible
for the breaking of the conformal symmetry~\footnote{Other (different)
aspects of this possibility have been considered in
Ref.~\cite{Higgs}.} and the generation of neutrino masses.  To see at
which scale this happens notice that conformal dynamics will still be
governing the right-handed neutrino sector until the mass becomes of
the order of the renormalization group scale. This happens when
\be
\mu_c= B^{1/2}_\gamma\,\frac{v}{\sqrt{2}}\left(\frac{\mu_c}{\Lambda}\right)^{\gamma}\,,
\ee  
and hence the physical Dirac mass will be 
\begin{equation}
m^D_\nu=\mu_c=v\,\left( \frac{B_\gamma}{2}\right)^{\frac{1}{2(1-\gamma)}}\left( \frac{v}{\Lambda} 
\right)^{\frac{\gamma}{1-\gamma}}\,.
\label{dirac}
\end{equation}
For a given neutrino mass Eq.~(\ref{dirac}) gives a relation between
the scale $\Lambda$ and the anomalous dimension $\gamma$ of the right
handed neutrino.  A plot of $\gamma$ as a function of $\Lambda$ for
various values of $m^D_\nu$ is given in Fig.~\ref{fig1}. It is very
interesting to observe that as $\Lambda$ is lowered from the GUT to
the TeV scale, the anomalous dimension roughly varies from
$\gamma=1/2$ [i.e.~the critical value at which a Majorana mass term
becomes an irrelevant perturbation, see Eq.~(\ref{Majoranamass})] to
$\gamma=1$ [the value at which the propagator becomes UV sensitive.]

Next consider the case of three generations. The best experimental
values for mass squared differences are $m_{2}^2-m_{1}^2\simeq 7.6\times
10^{-5}\, {\rm eV}^2$~\cite{Abe:2008ee} and $|m_{3}^2-m_2^2| \simeq
2.7\times 10^{-3}\, {\rm eV}^2$~\cite{Michael:2006rx}. In the case of a
regular hierarchy, $m_3> m_2> m_1$, this implies the bound on the
hierarchy
\be
%r_{32}\equiv
\frac{m_3}{m_2}\lsim 6\,,
\ee
while $m_2/m_1$ can be any number greater than one. Conversely, for
the case of an inverted hierarchy, $m_2>m_1> m_3$, one has
$m_2/m_1\approx 1$ while $m_2/m_3$ is unconstrained from above.  Let
us now introduce three right-handed neutrinos.
%~\footnote{In principle less
%than three right-handed neutrinos are possible but will lead to at
%least one vanishing eigenvalue.}
 Several possibilities can be proposed to reproduce the little
neutrino hierarchy.
\begin{itemize}
\item
The first possibility is having all the three right-handed neutrinos
with identical anomalous dimension.  In either of the two hierachy
schemes, the ratio of the two heavier masses is close to unity, and
the small splitting will be accounted for by SM corrections to the
Yukawa couplings. The same holds true if the lightest neutrino mass is
of the same order.  
\item
A much lighter state can also be naturally achieved. Let us work in
the regular hierarchy scheme for definiteness
and assign a larger anomalous dimension $\gamma_1$ to one of the right
handed neutrinos while keeping the other two equal,
e.g.~$\gamma_2=\gamma_3=\gamma$. At the scale $\mu_c$ the two heavy
neutrinos decouple and the running mass for the third neutrino equals
\be
m_1(\mu_c)
=\epsilon\, \mu_c \ll\mu_c\,,\qquad
\epsilon
=\left(\frac{\mu_c}{\Lambda}\right)^{\gamma_1-\gamma}\ll 1\,.
\label{bc}
\ee
However below $\mu_c$ the strongly coupled sector will flow away from
the fixed point. Without making further assumptions on that sector we
do not know which value $m_1$ will flow to in the IR.
\item
Assuming that the right-handed neutrino sector becomes weakly coupled
at the scale $\mu_c$, the physical mass $m_1$ will be given by
Eq.~(\ref{bc}) to good approximation.
\item
On the other hand, assuming that the flow below $\mu_c$ is governed by
a different IR fixed point with an anomalous dimension $\gamma_1'$ for
the remaining neutrino, we compute
\be
\mu_c'=m_1=\epsilon^\frac{1}{1-\gamma_1'} \mu_c \ll m_2,m_3\,,
\ee
leading to a further suppression of $m_1$ below $\mu_c$.  For this to
be efficient $\epsilon$ needs not even be particularly small. Instead
of generating it from a difference in the $\gamma$'s it can
just as well originate from a moderate Yukawa hierarchy at the high
scale, e.g.~$h_2=h_3\sim 1$ while $h_1\equiv \epsilon\sim 0.1$.
\item
Finally the last (obvious) possibility is that different neutrinos
$\nu_i$ belong to different conformal theories at the scale
$\Lambda_i$ and develop different anomalous dimensions $\gamma_i$
which can then describe different masses as in Fig.~\ref{fig1}.
\end{itemize}

\begin{figure}[htb]
\begin{minipage}{7.5cm}
\centering
\begin{picture}(100,62)(0,0)
\ArrowLine(0,0)(25,0)
\ArrowLine(25,0)(75,0)
\ArrowLine(75,0)(100,0)
\PhotonArc(50,0)(25,0,180){4}{6}
\Photon(50,25)(50,50){-4}{2}
\Text(12,-4)[tc]{$\mu$}
\Text(50,-4)[tc]{$\nu$}
\Text(87,-4)[tc]{$e$}
\Text(25,15)[br]{$W$}
\Text(75,15)[bl]{$W$}
\Text(55,50)[lt]{$\gamma$}
\end{picture} 
\\
\begin{picture}(100,50)(0,0)
\ArrowLine(0,0)(25,0)
\ArrowLine(25,0)(75,0)
\ArrowLine(75,0)(100,0)
\PhotonArc(50,0)(25,0,180){4}{6}
\Photon(12,0)(12,25){-4}{2}
\Text(12,-4)[tc]{$\mu$}
\Text(50,-4)[tc]{$\nu$}
\Text(87,-4)[tc]{$e$}
\Text(50,32)[bc]{$W$}
\Text(0,25)[tl]{$\gamma$}
\end{picture}
\begin{picture}(100,50)(0,0)
\ArrowLine(0,0)(25,0)
\ArrowLine(25,0)(75,0)
\ArrowLine(75,0)(100,0)
\PhotonArc(50,0)(25,0,180){4}{6}
\Photon(87,0)(87,25){-4}{2}
\Text(12,-4)[tc]{$\mu$}
\Text(50,-4)[tc]{$\nu$}
\Text(87,-4)[tc]{$e$}
\Text(50,32)[bc]{$W$}
\Text(100,25)[tr]{$\gamma$}
\end{picture}
\vspace{1cm}

\end{minipage}
\hfill
\begin{minipage}{7cm}

\psfrag{logLambda}{$\log_{10}(\Lambda/$GeV)}
\psfrag{logB}[][t]{$\log_{10}B(\mu\to e\gamma)$}
\includegraphics[width=7cm,height=5cm]{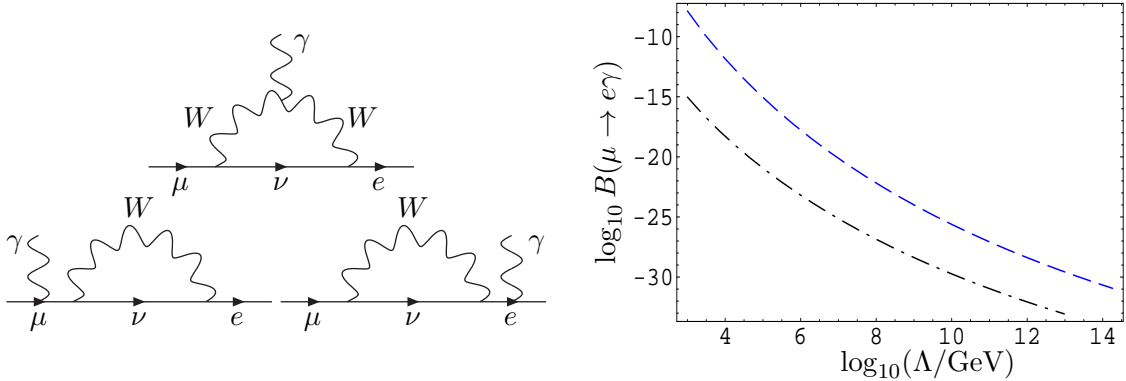}

\end{minipage}
\caption{Diagrams contributing to $\mu\to e\gamma$ (left panel) and the
branching ratio of that reaction as a function of the scale $\Lambda$
(right panel). The two lines correspond to $m_3=0.05$ eV (dashed blue line)
and $1$ eV (dash-dotted black line).}
\label{fcnc}
\end{figure}

Given that the Yukawa couplings and hence the Dirac masses grow with
energy one should be worried about possible flavor changing neutral
currents (FCNC). There are strong experimental bounds on decay
channels such as $\mu\to e\gamma$ and $\mu\to 3e$. We will see how
these relate to a bound on the anomalous dimension $\gamma$.  Focusing
on $\mu\to e \gamma$, one needs to evaluate the diagrams in the left
panel of Fig.~\ref{fcnc}.  The amplitudes actually go smoothly to zero
when the Dirac masses are turned off, so we will calculate their
effects perturbatively.  The leading contribution comes from two mass
insertions. All diagrams are IR and UV finite for $\gamma<1$, and the
amplitude can be parametrized as~\footnote{Under the assumption that
all external momenta are small compared to $M_W$.}
\be
\mathcal A \propto 
 \sum_i U^{\phantom *}_{ei}U^*_{\mu i}\, 
\frac{\pi\gamma\, B_\gamma\,(h_i v\Lambda^{-\gamma})^2}{\sin(\pi\gamma)M_W^{2-2\gamma}}\, 
= \frac{\pi\gamma}{\sin(\pi\gamma)}\sum_i U^{\phantom *}_{ei}U^*_{\mu i}\,
\left (\frac{m_i}{M_W}\right)^{2-2\gamma}\,.
\label{mutoegamma}
\ee
where we can see that the $\gamma$ dependence is a typical unparticle
effect since the right-handed unparticle propagates along the internal
lines of the diagram. Normalizing this to the main channel $\mu\to e
\nu\nu $ we get for the branching ratio
\be
B(\mu\to e \gamma)=\frac{3}{32}\frac{\alpha}{\pi}
\left | \frac{\pi\gamma}{\sin(\pi\gamma)}
\sum_i U^{\phantom *}_{ei}U^*_{\mu i}\,
\left (\frac{m_i}{M_W}\right)^{2-2\gamma}\right|^2\,.
\ee
The result for the SM with massive Dirac
neutrinos~\cite{Bilenky:1987ty}, recovered in the limit $\gamma\to 0$,
is known to be many orders of magnitudes below the experimental bound
$B(\mu\to e \gamma)< 1.2\times 10^{-11}$~\cite{Brooks:1999pu}. Due to
its exponential dependence on the anomalous dimension $B(\mu\to
e\gamma)$ can nevertheless reach this bound if $\gamma$ becomes close
enough to one.  Using the best fit values for
$U$~\cite{Schwetz:2008er}, the existing bound implies that
$\gamma\lsim 0.86$ for the case of a regular hierarchy with $m_1\ll
m_2,m_3$~\footnote{This does not change much for an inverted hierarchy
as long as $\theta_{13}$ is not too close to $\pi/2$. }. Future
experiments~\cite{Grassi:2005ac} aim to improve the sensitivity to
$\sim 10^{-14}$ which would push down the sensitivity to $\gamma\sim
0.81$. These bounds will be moving closer to $\gamma=1$ when the
neutrinos become more degenerate.  In Fig.~\ref{fcnc} we plot the
branching ratio as a function of the scale $\Lambda$.  We expect
similar bounds to hold from the $\mu\to 3 e$ channel as well as from
$\mu\to e$ conversion~\cite{prep}.

\begin{figure}[htb]\begin{center}
\psfrag{logLambda}{$\log_{10}(\Lambda/$GeV)}
\psfrag{B}{$\hspace{-.9cm} B(h\to\nu\bar\nu)$}
%\psfrag{d}[][b]{$d$}\psfrag{mpole}[][b]{$m_p$}
\psfrag{logh}[][t]{$\log_{10}h_\nu(v)$}
\psfrag{logML}[][t]{$\log_{10}(m_{\nu_L}^M/$eV)}
\includegraphics[width=7cm,height=5cm]{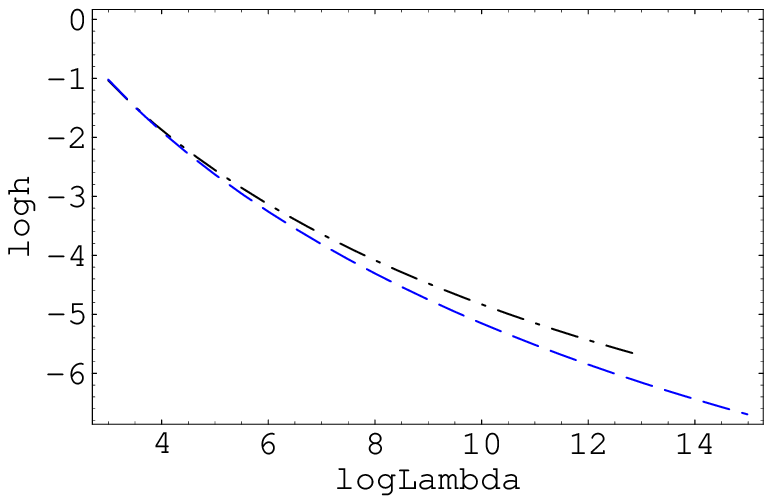}\hfill
\includegraphics[width=7cm,height=5cm]{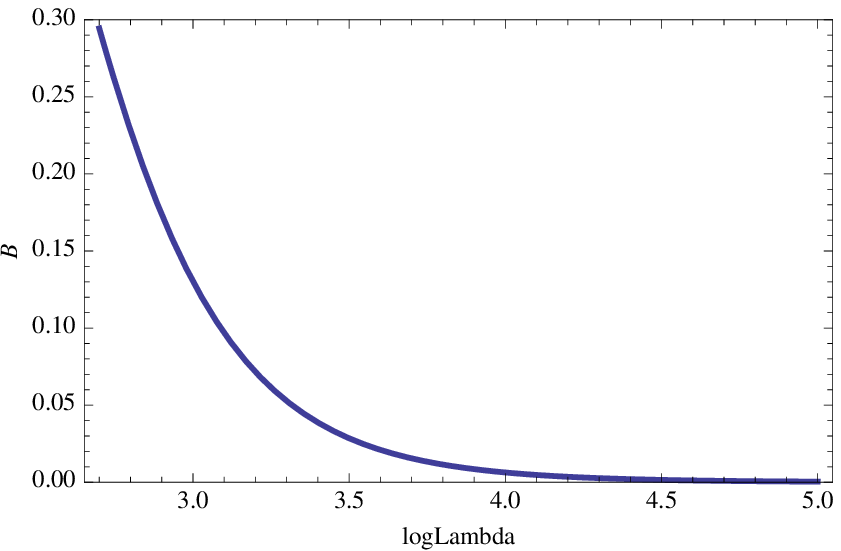}
\caption{Left panel: the Yukawa coupling at $\mu=v$ as a function of
the scale $\Lambda$. The two lines represent the same values for the
neutrino masses as in Fig.~\ref{fig1}. Right panel: the branching
ratio $B(h\to\nu\bar\nu)$ for $m_H=130$ GeV and three (almost)
degenerate neutrinos with mass $m_\nu^D\simeq 0.1$ eV as a function of
the scale $\Lambda$.}
\label{fig2}\end{center}
\end{figure}

Another very interesting effect that arises is that at a given scale
$\mu$ the neutrino Yukawa coupling is given by
\begin{equation}
h_\nu(\mu)=B_\gamma^{1/2}\,\left(\frac{\mu}{\Lambda}\right)^{\gamma}\,,
\label{acoplo}
\end{equation}
and it can be sizable at the LHC scales which are sensitive to the
electroweak scale. In the left panel of Fig.~\ref{fig2} the Yukawa
coupling is
plotted as a function of $\Lambda$ for the same values of the neutrino
masses as in Fig.~\ref{fig1} and $\mu=v$. From this one can see that
for a rather low cutoff e.g.~$\Lambda~\sim 10$ TeV, corresponding to
$\gamma\sim 0.8 - 0.9$, the neutrino Yukawa can be of the same order
as the $\tau$ or $c$ couplings, i.e.~at the percent level. In such an
extreme case one can even hope to have a sizable fraction of Higgs
decaying to neutrinos at the LHC. In fact one can easily compute the
width $\Gamma(h\to\nu\bar\nu)$ as the imaginary part of the one-loop
correction to the Higgs inverse propagator with a neutrino-loop
internal line and using the unparticle right-handed neutrino
propagator given in Eq.~(\ref{prop}). The result is given by
\begin{equation}
\Gamma(h\to\nu\bar\nu)=h_\nu^2(m_H)\,\frac{m_H}{16\pi}\,\frac{2}{\Gamma(1-\gamma)\Gamma(3+\gamma)}
\label{width}
\end{equation}
where $m_H$ is the Higgs pole mass and $h_\nu(m_H)$ is the neutrino
Yukawa coupling defined in Eq.~(\ref{acoplo}) at the scale
$\mu=m_H$. In the particle limit $\gamma\to 0$ the last factor in
(\ref{width}) goes to one and one recovers the Standard Model
expression
\begin{equation}
\Gamma_{\rm SM}(h\to\nu\bar\nu)=h_{\nu\,{\rm SM}}^2\,\frac{m_H}{16\pi}
\label{widthSM}
\end{equation}
By comparison of (\ref{width}) and (\ref{widthSM}) one can see that
the main difference between both expressions is the "conformal
running" of the neutrino Yukawa coupling. In the right panel of
Fig.~\ref{fig2} we plot the branching ratio with respect to the
dominant decay mode $h\to b\bar b$
\begin{equation}
B(h\to\nu\bar\nu)=\frac{\sum_i\Gamma(h\to\nu_i\bar\nu_i)}{\sum_i\Gamma(h\to\nu_i\bar\nu_i)+\Gamma(h\to b\bar b)}
\end{equation}
for the value of the Higgs mass $m_H=130$ GeV and corresponding to
three neutrino flavors (quasi) degenerate in mass. One can see that
for $\Lambda\lsim 10$ TeV the branching ratio corresponding to the
three neutrino channel is comparable (or dominant) to the branching
ratio into $c\bar c$ and $\tau\bar\tau$, which might have implications
for light Higgs searches.  More details will be given
elsewhere~\cite{prep}.

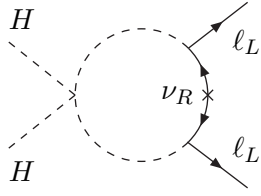
\begin{figure}
\centering
\begin{picture}(90,70)(0,0)
\DashCArc(50,35)(25,45,-45){3}
\ArrowArc(50,35)(25,0,45)
\ArrowArcn(50,35)(25,0,-45)
%\CArc(50,35)(25,-45,45)
\Line(73,33)(77,37)
\Line(77,33)(73,37)
\ArrowLine(67.7,52.7)(90,70)
\ArrowLine(67.3,17.3)(90,0)
\DashLine(25,35)(0,55){3}
\DashLine(25,35)(0,15){3}
\Text(0,10)[tl]{$H$}
\Text(0,60)[bl]{$H$}
\Text(90,60)[tc]{$\ell_L$}
\Text(90,10)[bc]{$\ell_L$}
\Text(70,35)[cr]{$\nu_R$}
\end{picture}
\caption{A diagram contributing to Eq.~(\ref{LV}).}
\label{diag}
\end{figure}

Let us finally comment on the possibility that the theory above the
scale $\Lambda$ violates lepton number. For values of $\gamma>1/2$ the
right handed Majorana mass operator
\be
\mathcal L^M=\frac{1}{2}\Lambda^{1-2\gamma}\mathcal N_R\mathcal N_R+h.c.\,,
\label{Majoranamass}
\ee
is an irrelevant perturbation and does not lead to a breakdown of the
conformal symmetry~\footnote{We assume here, for simplicity, that the
dimension of the mass operator is twice that of the neutrino field, as
it is the case for chiral fields in superconformal
theories~\cite{Seiberg}.}.  It can also be immediately verified that
at the scale $m^D_\nu$ the right-handed neutrino Majorana mass
$m^M_{\nu_R}(\mu)=B_\gamma\, (\mu/\Lambda)^{2\gamma}\Lambda$ is
parametrically suppressed with respect to the Dirac mass at the scale
of conformal breaking
\begin{equation}
m^M_{\nu_R}=\Lambda\left(
\frac{B_\gamma v^2}{2\Lambda^2}\right)^\frac{\gamma}{1-\gamma}
=m_\nu^D\, \left(\frac{m_\nu^D}{\Lambda}\right)^{2\gamma-1}.
\label{RH}
\end{equation}
However the fact that lepton number is violated also allows for
further higher dimension operators, the most important one being
\be
\mathcal L^{L\!\!\!/}= c\,\Lambda^{-1} (H\ell_L)^2\,.
\label{LV}
\ee
If this operator is generated from integrating out heavy right-handed
neutrinos at $\Lambda$ the constant $c$ will be of $\mathcal
O(1)$. This is of course the standard see-saw mechanism.  On the other
hand if the right-handed neutrinos are conformal (as we are assuming
in this paper) we cannot integrate them out.  Nevertheless due to the
presence of the irrelevant lepton-number violating operator $\mathcal
L^M$ loop corrections will still generate $\mathcal L^{L\!\!\!/}$. For
instance, the diagram in Fig.~\ref{diag} gives
\be
c(\gamma)=\frac{\lambda\, (B_\gamma)^2}{16\pi^2(2\gamma-1)}
   \left[1-\left(\frac{v}{\Lambda}\right)^{4\gamma-2}\right]
\,,
\ee
where $\lambda$ is the Higgs quartic coupling and 
one Majorana-mass insertion was used. Similar contributions will
be generated by box diagrams containing electroweak gauge bosons in
internal lines.  Once electroweak symmetry is broken a left-handed
neutrino mass
\be
m_{\nu_L}^M=c\, \frac{v^2}{\Lambda}\,,
\label{LH}
\ee
 is generated. For $\mu<v$ this mass will run much less compared to
the strongly running Dirac and right-handed Majorana masses and hence
the dominant neutrino mass will be
%~\footnote{There is one special
%case, $\gamma\simeq1/2$, in which $m_\nu^D\simeq v^2/\Lambda$ is actually
%comparable or larger than $m_{\nu_L}^M=c\, v^2/\Lambda$. }
$m_{\nu_L}^M$.  The neutrino mass matrix at the scale of conformal
breaking
\be
\mathcal M_\nu=
\left(
\begin{array}{cc}
m_{\nu_L}^M& m^D_\nu\\
m_\nu^D&m_{\nu_R}^M
\end{array}
\right)
\ee
has entries given by Eqs.~(\ref{dirac}), (\ref{RH}), and (\ref{LH})
respectively. Due to the hierarchy $m_{\nu_L}^M\gg m^D_\nu\gg
m_{\nu_R}^M$ we now have an {\em inverted} see-saw mechanism, with an
extremly light and almost completely sterile right-handed
neutrino~\footnote{A similar scenario has previously been proposed in
theories with extra dimensions~\cite{Xdim}.}.  The light mass eigenvalue
$m^{\prime M}_{\nu_L}$ and the mixing angle $\alpha$ are given by
\be
m^{\prime M}_{\nu_R}= m^M_{\nu_L}\, \alpha^2\,,\qquad \alpha=
   \left(\frac{B_\gamma^{1/2}}{\sqrt 2 c}\right)^{\frac{\gamma}{1-\gamma}}
   \left(\frac{m^M_{\nu_L}}{v}\right)^\frac{2\gamma-1}{1-\gamma}\,.
\ee
\begin{figure}[htb]\begin{center}
\psfrag{logLambda}{$\log_{10}(\Lambda/$GeV)}
\psfrag{g}[][t]{$\gamma$}
%\centering
\includegraphics[width=9cm,height=7.5cm]{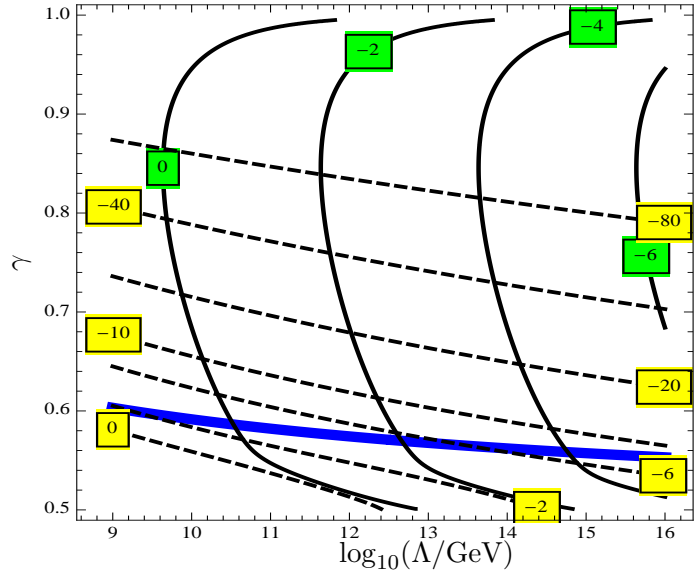}
\caption{Contour plots for fixed values of the mass of the heaviest
--mainly $\nu_L$-- (solid) and the lightest --mainly $\nu_R$--
(dashed) mass eigenstates.  The labels indicate the values of the
corresponding mass eigenvalues $m$ as $\log_{10}(m/{\rm eV})$. The region
below the thick (blue) line corresponds to mixing angles
$\alpha>0.1$.}
\label{ultima}\end{center}
\end{figure}

The light and heavy eigenvalues are displayed in
Fig.~\ref{ultima}. For $\gamma \simeq 1/2$ the mixing becomes of
$\mathcal O(1)$ and the two eigenvalues are similar. Such a scenario
would lead to modifications in the predictions for neutrino masses and
mixings as new sterile neutrinos participate in the
oscillations. Although mixing schemes with sterile neutrinos have been
proposed to explain the LSND anomaly~\cite{Peltoniemi:1993ec}, global
fits including recent data produce poor results~\cite{Maltoni:2007zf}. We therefore impose a rough cutoff of
$\alpha<0.1$ above which (the region below the thick line in Fig.~\ref{ultima}) we consider our model slightly
disfavoured.  It is thus interesting to notice that neutrino
oscillation experiments are able to probe our model even for large
values of $\Lambda$, as required in the lepton number violating
case~\footnote{We should point out here that it seems impossible to
expect a signal from Higgs decay for such large values of $\Lambda$,
as the Yukawa coupling at the weak scale will still be largely
suppressed. As for the $\mu\to e\gamma$ process the diagrams in
Fig.~\ref{fcnc} contain an internal left-handed neutrino line and now
there are no unparticles propagating in internal lines. The result
should then correspond to the usual one in the Standard Model
calculation (i.e.~that in Eq.~(\ref{mutoegamma}) for $\gamma\to 0$)
which is, for realistic neutrino masses, far away from the
experimental bounds.}. Needless to say in this case Majorana neutrino masses are
sensitive to neutrino-less double beta-decay experiments~\cite{Avignone:2007fu}. Finally, 
when lepton number is violated in the conformal sector, since the only 
$\gamma$-dependence for left-handed masses comes from the prefactor $c(\gamma)$,
it is natural to expect near-degenerate neutrino masses even for sizable differences in their
corresponding anomalous dimensions.

To conclude a conformally invariant right-handed neutrino sector
represents a natural way to obtain sub-eV neutrino masses if the
anomalous dimensions lie in the interval $1/2<\gamma<1$.  Electroweak
symmetry breaking triggers conformal breaking, which finally occurs at
the neutrino Dirac mass scale. The unusually strong energy dependence of the
right-handed neutrino field induces a series of interesting phenomena
which could be detected at future experiments.  Our model also
predicts lepton flavor violating reactions such as $\mu\to e\gamma $
at a much larger rate than in the Standard Model, and even opens up
the possibility to experimentally determine the anomalous dimensions
of the right-handed neutrinos in forthcoming experiments. Finally for
rather low scales $\Lambda\sim 10$ TeV the neutrino Yukawa couplings
can be comparable with those for charm and tau at the weak scale and
induce sizable (invisible) Higgs decay into the $\nu\bar\nu$ channel.
If the conformal theory violates lepton number small Majorana masses
can be generated without heavy states. Light sterile neutrinos then appear
and neutrino oscillation experiments are able to probe our model and put a
lower bound on the anomalous dimension of the right-handed
neutrino as $\gamma\gtsim 0.6$.

\subsection*{Acknowledgments}

\noindent 
Work supported in part by the European Commission under the European
Union through the Marie Curie Research and Training Networks ``Quest
for Unification" (MRTN-CT-2004-503369) and ``UniverseNet"
(MRTN-CT-2006-035863); by the Spanish Consolider-Ingenio 2010
Programme CPAN (CSD2007-00042); and by CICYT, Spain, under contract
FPA 2008-01430


\begin{thebibliography}{10}

\bibitem{Minkowski:1977sc}
  P.~Minkowski,
  %``Mu $\to$ E Gamma At A Rate Of One Out Of 1-Billion Muon Decays?,''
  Phys.\ Lett.\  B {\bf 67} (1977) 421;
  %%CITATION = PHLTA,B67,421;%%
  M.~Gell-Mann, P.~Ramond and R.~Slansky, in  {\it Supergravity}, 
  Ed. P. van Nieuwenhuizen and O. Freedman (North-Holland, Amsterdam, 1977);
  %
  T.~Yanagida, in {\it Proceedings of the Workshop on Unified Theories and 
  Baryon Number in the Universe}, Ed. A. Sawada and A. Sugamoto (KEK, Tsukoba, 1979);
%
%\bibitem{Mohapatra:1979ia}
  R.~N.~Mohapatra and G.~Senjanovic,
  %``Neutrino mass and spontaneous parity nonconservation,''
  Phys.\ Rev.\ Lett.\  {\bf 44} (1980) 912;
  %%CITATION = PRLTA,44,912;%%
%\bibitem{Mohapatra:1980yp}
%  R.~N.~Mohapatra and G.~Senjanovic,
  %``Neutrino Masses And Mixings In Gauge Models With Spontaneous Parity
  %Violation,''
  Phys.\ Rev.\  D {\bf 23} (1981) 165.
  %%CITATION = PHRVA,D23,165;%%


\bibitem{Georgi:1983mq}
  H.~Georgi, A.~E.~Nelson and A.~Manohar,
  %``On The Proposition That All Fermions Are Created Equal,''
  Phys.\ Lett.\  B {\bf 126} (1983) 169.
  %%CITATION = PHLTA,B126,169;%%

%\cite{Nelson:2000sn}
\bibitem{Nelson:2000sn}
  A.~E.~Nelson and M.~J.~Strassler,
  %``Suppressing flavor anarchy,''
  JHEP {\bf 0009} (2000) 030
  [arXiv:hep-ph/0006251].
  %%CITATION = JHEPA,0009,030;%%

\bibitem{Kobayashi:2001kz}
  T.~Kobayashi and H.~Terao,
  %``Sfermion masses in Nelson-Strassler type of models: SUSY standard  models
  %coupled with SCFTs,''
  Phys.\ Rev.\  D {\bf 64} (2001) 075003
  [arXiv:hep-ph/0103028].
  %%CITATION = PHRVA,D64,075003;%%


\bibitem{Nelson:2001mq}
  A.~E.~Nelson and M.~J.~Strassler,
  %``Exact results for supersymmetric renormalization and the supersymmetric
  %flavor problem,''
  JHEP {\bf 0207} (2002) 021
  [arXiv:hep-ph/0104051].
  %%CITATION = JHEPA,0207,021;%%

\bibitem{Luty:2001jh}
  M.~A.~Luty and R.~Sundrum,
  %``Supersymmetry breaking and composite extra dimensions,''
  Phys.\ Rev.\  D {\bf 65} (2002) 066004
  [arXiv:hep-th/0105137].
  %%CITATION = PHRVA,D65,066004;%%

\bibitem{Appelquist:2002me}
  T.~Appelquist and R.~Shrock,
  %``Neutrino masses in theories with dynamical electroweak symmetry
  %breaking,''
  Phys.\ Lett.\  B {\bf 548} (2002) 204
  [arXiv:hep-ph/0204141].
  %%CITATION = PHLTA,B548,204;%%


%\cite{Georgi:2007ek}
\bibitem{Georgi}
  H.~Georgi,
  %``Unparticle Physics,''
  Phys.\ Rev.\ Lett.\  {\bf 98} (2007) 221601
  [hep-ph/0703260]; 
%%CITATION = PRLTA,98,221601;%%
%\cite{Georgi:2007si}
%  H.~Georgi,
  %``Another Odd Thing About Unparticle Physics,''
  Phys.\ Lett.\  B {\bf 650}, 275 (2007)
  [hep-ph/0704.2457].
  %%CITATION = ARXIV:0704.2457;%%
%


\bibitem{Cacciapaglia:2008ns}
  G.~Cacciapaglia, G.~Marandella and J.~Terning,
  %``The AdS/CFT/Unparticle Correspondence,''
  arXiv:0804.0424 [hep-ph].
  %%CITATION = ARXIV:0804.0424;%%

\bibitem{Xdim}
  K.~R.~Dienes, E.~Dudas and T.~Gherghetta,
  %``Light neutrinos without heavy mass scales: A higher-dimensional seesaw
  %mechanism,''
  Nucl.\ Phys.\  B {\bf 557} (1999) 25
  [arXiv:hep-ph/9811428];
  %%CITATION = NUPHA,B557,25;%%
  N.~Arkani-Hamed, S.~Dimopoulos, G.~R.~Dvali and J.~March-Russell,
  %``Neutrino masses from large extra dimensions,''
  Phys.\ Rev.\  D {\bf 65} (2002) 024032
  [arXiv:hep-ph/9811448];
  %%CITATION = PHRVA,D65,024032;%%
  Y.~Grossman and M.~Neubert,
  %``Neutrino masses and mixings in non-factorizable geometry,''
  Phys.\ Lett.\  B {\bf 474} (2000) 361
  [arXiv:hep-ph/9912408];
  %%CITATION = PHLTA,B474,361;%%
  D.~Diego and M.~Quiros,
  %``Dirac Vs. Majorana Neutrino Masses From a TeV Interval,''
  Nucl.\ Phys.\  B {\bf 805} (2008) 148
  [arXiv:0804.2838 [hep-ph]].
  %%CITATION = NUPHA,B805,148;%%


\bibitem{Seiberg}
  K.~A.~Intriligator and N.~Seiberg,
  %``Lectures on supersymmetric gauge theories and electric-magnetic  duality,''
  Nucl.\ Phys.\ Proc.\ Suppl.\  {\bf 45BC} (1996) 1
  [arXiv:hep-th/9509066].
  %%CITATION = NUPHZ,45BC,1;%%

\bibitem{Higgs}
P.~J.~Fox, A.~Rajaraman and Y.~Shirman,
  %``Bounds on Unparticles from the Higgs Sector,''
  Phys.\ Rev.\  D {\bf 76} (2007) 075004
  [arXiv:0705.3092 [hep-ph]];
  %%CITATION = PHRVA,D76,075004;%%
%
 A.~Delgado, J.~R.~Espinosa and M.~Quiros,
  %``Unparticles-Higgs Interplay,''
  JHEP {\bf 0710} (2007) 094
  [arXiv:0707.4309 [hep-ph]];
  %%CITATION = JHEPA,0710,094;%%
%
 T.~Kikuchi and N.~Okada,
  %``Unparticle physics and Higgs phenomenology,''
  Phys.\ Lett.\  B {\bf 661} (2008) 360
  [arXiv:0707.0893 [hep-ph]].
  %%CITATION = PHLTA,B661,360;%%


\bibitem{Abe:2008ee}
  S.~Abe {\it et al.}  [KamLAND Collaboration],
  %``Precision Measurement of Neutrino Oscillation Parameters with KamLAND,''
  Phys.\ Rev.\ Lett.\  {\bf 100} (2008) 221803
  [arXiv:0801.4589 [hep-ex]];
  %%CITATION = PRLTA,100,221803;%%
%\cite{Ichimura:2008km}
%\bibitem{Ichimura:2008km}
  K.~Ichimura [KamLAND Collaboration],
  %``Recent Results from KamLAND,''
  arXiv:0810.3448 [hep-ex].
  %%CITATION = ARXIV:0810.3448;%%


%\cite{Michael:2006rx}
\bibitem{Michael:2006rx}
  D.~G.~Michael {\it et al.}  [MINOS Collaboration],
  %``Observation of muon neutrino disappearance with the MINOS detectors and the
  %NuMI neutrino beam,''
  Phys.\ Rev.\ Lett.\  {\bf 97} (2006) 191801
  [arXiv:hep-ex/0607088];
  %%CITATION = PRLTA,97,191801;%%
%\cite{Adamson:2007gu}
%\bibitem{Adamson:2007gu}
  P.~Adamson {\it et al.}  [MINOS Collaboration],
  %``A Study of Muon Neutrino Disappearance Using the Fermilab Main Injector
  %Neutrino Beam,''
  Phys.\ Rev.\  D {\bf 77} (2008) 072002
  [arXiv:0711.0769 [hep-ex]].
  %%CITATION = PHRVA,D77,072002;%%



\bibitem{Bilenky:1987ty}
  For a review see: S.~M.~Bilenky and S.~T.~Petcov,
  %``Massive Neutrinos and Neutrino Oscillations,''
  Rev.\ Mod.\ Phys.\  {\bf 59} (1987) 671
  [Erratum-ibid.\  {\bf 61} (1989\ ERRAT,60,575-575.1988) 169.1989\ ERRAT,60,575], and references therein.
  %%CITATION = RMPHA,59,671;%%


\bibitem{Brooks:1999pu}
  M.~L.~Brooks {\it et al.}  [MEGA Collaboration],
  %``New Limit for the Family-Number Non-conserving Decay mu+ to e+_gamma,''
  Phys.\ Rev.\ Lett.\  {\bf 83} (1999) 1521
  [arXiv:hep-ex/9905013].
  %%CITATION = PRLTA,83,1521;%%

\bibitem{Schwetz:2008er}
  T.~Schwetz, M.~Tortola and J.~W.~F.~Valle,
  %``Three-flavour neutrino oscillation update,''
  New J.\ Phys.\  {\bf 10} (2008) 113011
  [arXiv:0808.2016 [hep-ph]].
  %%CITATION = NJOPF,10,113011;%%



\bibitem{Grassi:2005ac}
  M.~Grassi  [MEG Collaboration],
  %``The Meg Experiment At Psi: Status And Prospect,''
  Nucl.\ Phys.\ Proc.\ Suppl.\  {\bf 149} (2005) 369.
  %%CITATION = NUPHZ,149,369;%%


\bibitem{prep}
G.~v.~Gersdorff and M.~Quir\'os, in preparation.

\bibitem{Amsler:2008zz}
  C.~Amsler {\it et al.}  [Particle Data Group],
  %``Review of particle physics,''
  Phys.\ Lett.\  B {\bf 667} (2008) 1.
  %%CITATION = PHLTA,B667,1;%%

\bibitem{Peltoniemi:1993ec}
  J.~T.~Peltoniemi and J.~W.~F.~Valle,
  %``Reconciling dark matter, solar and atmospheric neutrinos,''
  Nucl.\ Phys.\  B {\bf 406} (1993) 409
  [arXiv:hep-ph/9302316];
  %%CITATION = NUPHA,B406,409;%%
%bibitem{Caldwell:1993kn}
  D.~O.~Caldwell and R.~N.~Mohapatra,
  %``Neutrino mass explanations of solar and atmospheric neutrino deficits  and
  %hot dark matter,''
  Phys.\ Rev.\  D {\bf 48} (1993) 3259.
  %%CITATION = PHRVA,D48,3259;%%


\bibitem{Maltoni:2007zf}
  M.~Maltoni and T.~Schwetz,
  %``Sterile neutrino oscillations after first MiniBooNE results,''
  Phys.\ Rev.\  D {\bf 76} (2007) 093005
  [arXiv:0705.0107 [hep-ph]].
  %%CITATION = PHRVA,D76,093005;%%

\bibitem{Avignone:2007fu}
  F.~T.~Avignone, S.~R.~Elliott and J.~Engel,
  %``Double Beta Decay, Majorana Neutrinos, and Neutrino Mass,''
  Rev.\ Mod.\ Phys.\  {\bf 80} (2008) 481
  [arXiv:0708.1033 [nucl-ex]].
  %%CITATION = RMPHA,80,481;%%


\end{thebibliography}
\end{document}